\documentclass[aps,prd,preprint,tightenlines,nofootinbib,showpacs]{revtex4}

\usepackage{graphicx}
\usepackage{dcolumn}
\usepackage{bm}

\begin{document}

\preprint{CALT-68-2445}

\title{\boldmath Constraints on the CKM Angle $\gamma$ from $B\to K^{*\pm}\pi^\mp$}

\author{Werner M.~Sun}
\altaffiliation[Present address: ]
{Cornell University, Ithaca, New York 14853, USA}
\email[email: ]{wsun@mail.lepp.cornell.edu}
\affiliation{California Institute of Technology, Pasadena,
California 91125, USA}

\date{August 12, 2003}

\begin{abstract} 
We present constraints on the CKM parameter $\gamma=\arg V_{ub}^*$ formed
within the framework of SU(3) symmetry and 
based on charmless hadronic $B$ decays to $K^{*\pm}\pi^\mp$ and other
pseudoscalar-vector final states.
For strong phases of ${\cal O}(10^\circ)$, our analysis weakly favors
$\cos\gamma<0$.  We also estimate that a determination of $\gamma$ with an
experimental uncertainty of less than $10^\circ$ can be
attained with an order-of-magnitude improvement in the precision of the
experimental inputs, but SU(3) symmetry breaking could introduce corrections
approaching the size of the current experimental uncertainties.
\end{abstract}

\pacs{13.20.He}
\maketitle

In the Standard Model, the Cabibbo-Kobayashi-Maskawa (CKM) quark mixing
matrix~\cite{Kobayashi:fv} gives rise to $CP$-violating phenomena through
its single complex phase.  This phase can be probed experimentally
by measuring decay rates and $CP$ asymmetries for charmless hadronic $B$
decays that receive contributions from amplitudes with differing weak phases.
In the flavor SU(3) decomposition of the amplitudes in
pseudoscalar-vector ($PV$) final states~\cite{Dighe:1997wj},
the $b\to u\bar u s$ transition $B\to K^{*\pm}\pi^\mp$ is dominated by two
amplitudes, color-allowed tree and gluonic penguin, which interfere with a
weak phase $\pi-\gamma$, where $\gamma = \arg V_{ub}^*$, and with an unknown
strong phase $\delta$.  We extract $\cos\gamma$ in two ways, employing Monte
Carlo simulation to propagate experimental uncertainties and ratios of meson
decay constants to account for SU(3) symmetry breaking.  The first uses
only $B\to K^{*\pm}\pi^\mp$, and the second adds information from
$B\to\phi ^{^{(}}\!\bar K^{^{)}}$.  In both cases,
the magnitudes of the penguin and tree amplitudes must be known, and we
estimate these from CKM unitarity and measured branching fractions for other
$B\to PV$ decays.

An alternative method of constraining $\gamma$~\cite{Hocker:2001xe} makes use
of observables pertaining to $b\to c$ transitions and to mixing in the
neutral $B$ and $K$ systems, with a resultant 95\% confidence level (C.L.)
allowed interval for $\gamma$ of
$[38^\circ,80^\circ]$.  In contrast, the analysis presented in this
Letter uses rare, charmless $b\to u, d, s$
transitions without reference to mixing-induced $CP$ violation.
A discrepancy between the constraints on $\gamma$ from
charmless hadronic $B$ decays and those from $B$ and $K$ mixing
might arise from new physics contributions
to either $B$ and $K$ mixing or the $b\to s$ or $b\to d$ penguins.

Global analyses of charmless hadronic $B$ decays in the framework of
QCD-improved factorization~\cite{Du:2002cf,Aleksan:2003qi} find a value for
$\gamma$ of approximately $80^\circ$.  However, these fits predict smaller
branching fractions for $B\to K^{*\pm}\pi^\mp$ and
$^{^{(}}\!\bar K^{^{)}*0}\pi^\pm$ than are observed experimentally,
and removing these modes from the above analyses improves the fit quality.
It has been suggested~\cite{Keum:2002qj} that the $B\to K^*\pi$ modes may
receive dynamical enhancements not accounted for in
Refs.~\cite{Du:2002cf} and~\cite{Aleksan:2003qi}.  Our analysis focuses on
$B\to K^{*\pm}\pi^\mp$ with input from a modest number of other $B\to PV$
decays, thus providing a complement to the global fits.

Following the notation in Ref.~\cite{Dighe:1997wj} for SU(3) invariant
amplitudes, we denote color-allowed
tree amplitudes by $t$ and gluonic penguins by $p$.  Amplitudes for
$\left|\Delta S\right|=1$ transitions are primed, while those for
$\Delta S=0$ transitions are unprimed.  A subscript $P$ or $V$
indicates whether the spectator quark hadronizes into the
pseudoscalar or vector meson, respectively.
Since $B\to K^{*\pm}K^\mp$ is dominated by penguin annihilation and
$W$-exchange contributions or rescattering effects, and these decays have not
been observed experimentally, we neglect such amplitudes in our analysis.
The transition amplitude for $B\to K^{*\pm}\pi^\mp$ is
$A(K^{*\pm}\pi^\mp) = -(p_P' + t_P')$.  The amplitudes
$t'$ and $p'$ carry the CKM matrix elements $V_{ub}^* V_{us}$ and
$V_{tb}^* V_{ts}$ with weak phases $\gamma$ and $\pi$, respectively.
The amplitudes for the two charge states are given by
\begin{eqnarray}
A(K^{*+}\pi^-)&=&\left|p_P'\right|-\left|t_P'\right| e^{i\gamma} e^{i\delta} \\
A(K^{*-}\pi^+)&=&\left|p_P'\right|-\left|t_P'\right| e^{-i\gamma} e^{i\delta},
\end{eqnarray}
and we can express the $CP$-averaged amplitude as
\begin{equation}
\frac{1}{2}\left[\left|A(K^{*+}\pi^-)\right|^2 +\left|A(K^{*-}\pi^+)\right|^2 \right] =\left|p_P'\right|^2 + \left|t_P'\right|^2 - 2\left|p_P'\right|\left|t_P'\right|\cos\gamma\cos\delta.
\end{equation}

We identify squared amplitudes, $\left|A\right|^2=A^*A$, with branching
fractions, ${\cal B}$, and
we absorb all numerical factors, like $G_F$, $m_B$, phase
space integrals, decay constants, form factors, and CKM matrix
elements, into the definitions of the amplitudes.  Most of the $B$
branching fraction measurements in the literature are calculated
assuming equal production of charged and neutral mesons.  We correct these
branching fractions by the ratio of $B^+B^-$ to $B^0\bar B^0$ production rates,
$f_{+-}/f_{00}$, as well as by the ratio of charged to neutral lifetimes,
$\tau_+/\tau_0$.  Because the constraints on $\gamma$ are constructed from
ratios of branching fractions, we scale only the neutral $B$ branching
fractions by the product
${\cal F}\equiv\frac{f_{+-}}{f_{00}}\cdot\frac{\tau_+}{\tau_0}$, which is
measured directly in Refs.~\cite{Alexander:2000tb} and~\cite{Athar:2002mr}.
Thus, $\cos\gamma\cos\delta$ can be expressed in terms of the $CP$-averaged
branching fraction ${\cal B}(K^{*\pm}\pi^\mp)$:
\begin{equation}\label{eq:cosGammaCosDelta}
\cos\gamma\cos\delta = \frac{ \left|p_P'\right|^2 + \left|t_P'\right|^2 -
{\cal B}(K^{*\pm}\pi^\mp){\cal F} }
{ 2\left|p_P'\right|\left|t_P'\right| }.
\end{equation}
For a given value of $\delta$, $\gamma$ is determined to a twofold ambiguity.
The rate difference between $\bar B^0\to K^{*-}\pi^+$ and
$B^0\to K^{*+}\pi^-$, which is proportional to $\sin\gamma\sin\delta$,
provides an additional observable
that allows us to disentangle $\gamma$ and $\delta$:
\begin{eqnarray}
\label{eq:cosGammaPlusDelta}
\cos(\gamma+\delta) &=& \frac{\left|p_P'\right|^2 + \left|t_P'\right|^2 -
	{\cal B}(K^{*+}\pi^-){\cal F} }{2\left|p_P'\right| \left|t_P'\right|}\\
\label{eq:cosGammaMinusDelta}
\cos(\gamma-\delta) &=& \frac{\left|p_P'\right|^2 + \left|t_P'\right|^2 -
	{\cal B}(K^{*-}\pi^+){\cal F} }{2\left|p_P'\right| \left|t_P'\right|},
\end{eqnarray}
which leads to
\begin{eqnarray}
\label{eq:gamma}
\gamma &=& \frac{1}{2}\left[
\cos^{-1}\frac{\left|p_P'\right|^2 + \left|t_P'\right|^2 -
	{\cal B}(K^{*+}\pi^-){\cal F} }
{2\left|p_P'\right| \left|t_P'\right|} +
\cos^{-1}\frac{\left|p_P'\right|^2 + \left|t_P'\right|^2 -
	{\cal B}(K^{*-}\pi^+){\cal F} }
{2\left|p_P'\right| \left|t_P'\right|}
\right] \\
\label{eq:delta}
\delta &=& \frac{1}{2}\left[
\cos^{-1}\frac{\left|p_P'\right|^2 + \left|t_P'\right|^2 -
	{\cal B}(K^{*+}\pi^-){\cal F} }
{2\left|p_P'\right| \left|t_P'\right|} -
\cos^{-1}\frac{\left|p_P'\right|^2 + \left|t_P'\right|^2 -
	{\cal B}(K^{*-}\pi^+){\cal F} }
{2\left|p_P'\right| \left|t_P'\right|}
\right].
\end{eqnarray}
These expressions for $\gamma$ and $\delta$ are subject to a
fourfold ambiguity: $\{\gamma,\delta\}\to\{\delta,\gamma\}$,
$\{-\gamma,-\delta\}$, or $\{-\delta,-\gamma\}$.

The charge-separated branching fractions ${\cal B}(K^{*+}\pi^-)$ and
${\cal B}(K^{*-}\pi^+)$ appearing in
Eqs.~\ref{eq:cosGammaPlusDelta}--\ref{eq:delta} can be determined directly
from ${\cal B}(K^{*\pm}\pi^\mp)$ and the $CP$ asymmetry
${\cal A}_{CP}(K^{*\pm}\pi^\mp)$.
In addition, SU(3) symmetry relates the rate difference
$\Delta(K^{*-}\pi^+)\equiv{\cal B}(K^{*-}\pi^+)-{\cal B}(K^{*+}\pi^-)$ to
the corresponding $\Delta S=0$ quantity,
$\Delta(\rho^-\pi^+)\equiv{\cal B}(\bar B^0\to\rho^-\pi^+)-{\cal B}(B^0\to\rho^+\pi^-)$~\cite{Deshpande:2000jp,Dariescu:2002hw}:
\begin{equation}\label{eq:deltaRhopi}
\Delta(K^{*-}\pi^+) =
        -\left[\frac{f_{K^*}F_1^{B\to\pi}(m_{K^*}^2)}
        {f_\rho F_1^{B\to\pi}(m_\rho^2)}\right]^2
        \Delta(\rho^-\pi^+).
\end{equation}
To attain greater precision on ${\cal B}(K^{*+}\pi^-)$ and
${\cal B}(K^{*-}\pi^+)$, we combine information on $\Delta(\rho^-\pi^+)$ with
the measurements of ${\cal B}(K^{*\pm}\pi^\mp)$ and
${\cal A}_{CP}(K^{*\pm}\pi^\mp)$ listed in Table~\ref{table:inputSummary}.
These inputs are given relative weights that minimize the
uncertainties on ${\cal B}(K^{*+}\pi^-)$ and
${\cal B}(K^{*-}\pi^+)$, and we account for the correlation between
CLEO's ${\cal A}_{CP}(K^{*\pm}\pi^\mp)$ and ${\cal B}(K^{*\pm}\pi^\mp)$
measurements, which are made with the same dataset and technique.

The {\sc BaBar} analysis of $B\to\pi^+\pi^-\pi^0$~\cite{Babar3pi},
which determines the $CP$ asymmetry and dilution parameters $A$, $C$, and
$\Delta C$ defined in Ref.~\cite{Babar3pi}, allows us to evaluate
$\Delta(\rho^-\pi^+)=-(A+C+A\Delta C)\cdot{\cal B}(\rho^\pm\pi^\mp)$.
We propagate the uncertainties on these parameters with their
correlations~\cite{Hoecker} to obtain
$\Delta(\rho^-\pi^+)=-(2.9\pm 4.6)\cdot 10^{-6}$.
Thus, taking the form factor ratio in Eq.~\ref{eq:deltaRhopi} to be unity,
we find
${\cal B}(K^{*+}\pi^-) = (14.4^{+4.4}_{-4.0})\cdot 10^{-6}$ and
${\cal B}(K^{*-}\pi^+) = (18.7^{+4.8}_{-4.6})\cdot 10^{-6}$
with a correlation coefficient of $0.53$.  The correlation coefficients
between $\Delta(\rho^-\pi^+)$ and these two branching fractions are
$0.45$ for ${\cal B}(K^{*+}\pi^-)$ and $-0.50$ for ${\cal B}(K^{*-}\pi^+)$.

A second method of estimating $\gamma$ uses
$B\to K^{*\pm}\pi^\mp$ and $B\to\phi K^\pm$.  The possibility of
constraining $\gamma$ from these decays was first noticed by
Gronau and Rosner~\cite{gronau-rosner}, and the concrete formulation of this
method was subsequently put forth by Gronau~\cite{Gronau:az,gronau2}.
The SU(3) decomposition of the $B\to\phi K^\pm$ amplitude is
$A(\phi K^\pm ) = p_P'-\frac{1}{3} P_{\rm EW}^{\prime P}$ to
${\cal O}(\lambda)$, where $P_{\rm EW}^{\prime P}$
denotes the electroweak penguin contribution.
The weak phase of $P_{\rm EW}^{\prime P}$ is the same as that of $p_P'$, and
its strong phase is expected to be the same as in $t_P'$ because of the
similarity of their flavor topologies~\cite{Gronau:az,gronau2}.
Thus, the ratio of the $CP$-averaged branching fractions for
$B\to K^{*\pm}\pi^\mp$ and $B\to\phi K^\pm$
provides a measure of $\gamma$ up to a twofold ambiguity:
\begin{equation}\label{eq:cosGammaPhik}
\cos\gamma = \frac{1}{2r\cos\delta}
\left[ 1 + r^2 - R \left( 1-
2\cos\delta\left|\frac{P_{\rm EW}^{\prime P}}{3p_P'}\right|+
\left|\frac{P_{\rm EW}^{\prime P}}{3p_P'}\right|^2 \right)\right],
\end{equation}
where $r\equiv \left|t_P'/p_P'\right|$, and
\begin{equation}\label{eq:cosGammaPhik2}
R \equiv
\frac{\left|A(K^{*+}\pi^-)\right|^2 + \left|A(K^{*-}\pi^+)\right|^2}
{\left|A(\phi K^+)\right|^2 + \left|A(\phi K^-)\right|^2}.
\end{equation}
Both $B\to\phi K^\pm$ and
$B\to\phi ^{^{(}}\!\bar K^{^{)}0}$ receive the same SU(3)
amplitude contributions~\cite{Dighe:1997wj}, so we can improve the statistical
precision of Eq.~\ref{eq:cosGammaPhik2} by combining both channels:
\begin{equation}
R= \frac{{\cal B}(K^{*\pm}\pi^\mp){\cal F}}
{\left[\sigma_0^2 {\cal B}(\phi K^\pm) +
\sigma_+^2 {\cal B}(\phi ^{^{(}}\!\bar K^{^{)}0}){\cal F}
\right]/
(\sigma_+^2 + \sigma_0^2)},
\end{equation}
where $\sigma_+$ and $\sigma_0$ refer to the uncertainties
on ${\cal B}(\phi K^\pm)$ and
${\cal B}(\phi ^{^{(}}\!\bar K^{^{)}0}){\cal F}$,
respectively.  To determine $\gamma$ with this method, the size of $\delta$
must be known.  It is believed, based on perturbative~\cite{delta1} and
statistical~\cite{delta2} calculations, that
$0^\circ < \left|\delta\right| < 90^\circ$.  In the simulation, we fix
$|P_{\rm EW}^{\prime P}|$ to be $\frac{1}{2}\left|p_P'\right|$,
as given by factorization
calculations~\cite{gronau-rosner,fleischer,deshpande-he}, and we evaluate the
dependence of our results on $|P_{\rm EW}^{\prime P}/p_P'|$ and
$\delta$.

In both of the above methods of constraining $\gamma$
(involving Eqs.~\ref{eq:cosGammaCosDelta}--\ref{eq:delta} and
Eq.~\ref{eq:cosGammaPhik}), numerical values of $\left|t_P'\right|$ and
$\left|p_P'\right|$ are given by other $B\to PV$
branching fractions~\cite{Chiang:2001ir}.  The penguin amplitude is simply
\begin{equation}
\left|p_P'\right| = \sqrt{{\cal B}(^{^{(}}\!\bar K^{^{)}*0}\pi^\pm)}.
\end{equation}
The tree amplitude is taken from the $\Delta S = 0$
transition $B\to\rho^\pm\pi^\mp$ and related to the $\left|\Delta S\right|=1$
amplitude through SU(3)-breaking factors:
\begin{equation}
\left|t_P'\right| = \left|\frac{V_{us}}{V_{ud}}\right|
\frac{f_{K^*}}{f_\rho} \left|t_P\right|.
\end{equation}
The experimentally measured ${\cal B}(\rho^\pm\pi^\mp)$ represents a sum over
$B^0\to\rho^\pm\pi^\mp$ and $\bar B^0\to\rho^\pm\pi^\mp$ decays:
\begin{equation}
{\cal B}(\rho^\pm\pi^\mp) =
\frac{1}{\cal F}\left(\left|t_P+p_P\right|^2 + \left|t_V+p_V\right|^2\right).
\end{equation}
We isolate $\left|t_P+p_P\right|$ with the {\sc BaBar} $B\to\pi^+\pi^-\pi^0$
analysis~\cite{Babar3pi}, which provides
\begin{eqnarray}
{\cal B}(\rho^\pm\pi^\mp)_P &\equiv&
\frac{1}{2}\left[{\cal B}(B^0\to\rho^+\pi^-)+
{\cal B}(\bar B^0\to\rho^-\pi^+)\right] =
\frac{1}{\cal F}\left|t_P+p_P\right|^2 \\
&=& \frac{1}{2}{\cal B}(\rho^\pm\pi^\mp)\cdot(1+AC+\Delta C).
\end{eqnarray}
Based on the experimental inputs in Table~\ref{table:inputSummary}, we find
${\cal B}(\rho^\pm\pi^\mp)_P=(13.9\pm 2.7)\cdot 10^{-6}$
and a correlation coefficient between ${\cal B}(\rho^\pm\pi^\mp)_P$ and
$\Delta(\rho^-\pi^+)$ of 0.05.

Extracting $\left|t_P\right|$ from ${\cal B}(\rho^\pm\pi^\mp)_P$ requires
estimates of the magnitude and phase of $p_P$.  Its magnitude is obtained
from the analogous $\left|\Delta S\right|=1$ amplitude:
\begin{equation}
\left|p_P\right| = \left|\frac{V_{td}}{V_{ts}}\right|
\frac{f_\rho}{f_{K^*}} \left|p_P'\right|.
\end{equation}
In the SU(3) limit, $p_P$ and $t_P$ have the same relative strong phase
as that between $p_P'$ and $t_P'$.  Their relative weak phase, however,
is $\gamma+\beta$, where $\gamma$ is unknown, {\it a priori}.  Therefore, we must
solve for $\cos\gamma$ and $\left|t_P'\right|$ simultaneously.

Using CKM unitarity, the parameters $\left| V_{td}/V_{ts}\right|$ and $\beta$
can be eliminated in favor of $\left|V_{ub}/V_{cb}\right|$ and $\gamma$ via the
relations
\begin{eqnarray}
\left|\frac{V_{td}}{V_{ts}}\right|^2&=&
	\left|V_{us}\right|^2 -
	2\left|V_{us}\right|\left|\frac{V_{ub}}{V_{cb}}\right|\cos\gamma +
	\left|\frac{V_{ub}}{V_{cb}}\right|^2 \\
\sin\beta&=&\left|\frac{V_{ts}}{V_{td}}\right|
	\left|\frac{V_{ub}}{V_{cb}}\right|\sin\gamma\\
\cos\beta&=&\left|\frac{V_{ts}}{V_{td}}\right|
	\left(\left|V_{us}\right| -
	\left|\frac{V_{ub}}{V_{cb}}\right|\cos\gamma\right).
\end{eqnarray}
By making these substitutions, we remove our dependence
on $\sin 2\beta$ measurements involving $b\to c$
transitions and $B^0$-$\bar B^0$ mixing, and we remain sensitive to
new physics which may affect these processes and
charmless $b\to u,d,s$ transitions differently.

From the above unitarity relations and the $CP$-averaged branching fraction
\begin{equation}
{\cal B}(\rho^\pm\pi^\mp)_P = \left|p_P\right|^2 + \left|t_P\right|^2 +
2\left|p_P\right|\left|t_P\right|\cos(\gamma+\beta)\cos\delta,
\end{equation}
we find the following expression for $\left|t_P'\right|$:
\begin{equation}\label{eq:tpprime}
\left|t_P'\right|=\left|\frac{V_{us}}{V_{ud}}\right| \left|p_P'\right| y
\left\{ 1 \pm \sqrt{ 1 - \frac{1}{y^2} \left( \left|V_{us}\right|^2 -
2\left|V_{us}\right|\left|\frac{V_{ub}}{V_{cb}}\right|\cos\gamma +
\left|\frac{V_{ub}}{V_{cb}}\right|^2 \right) +
\frac{f_{K^*}^2{\cal B}(\rho^\pm\pi^\mp)_P{\cal F}}
{f_\rho^2\left|p_P'\right|^2 y^2}} \right\},
\end{equation}
where
\begin{equation}\label{eq:y}
y\equiv \left(\left|\frac{V_{ub}}{V_{cb}}\right|-
\left|V_{us}\right|\cos\gamma\right)\cos\delta.
\end{equation}
Using Eq.~\ref{eq:tpprime} to calculate $\left|t_P'\right|$ from
$B\to K^{*\pm}\pi^\mp$,
$^{^{(}}\!\bar K^{^{)}*0}\pi^\pm$, and
$\rho^\pm\pi^\mp$ depends on a choice of $\delta$ as well as
knowledge of $\gamma$, and an iterative solution is required.
The fixed strong phase appearing in Eq.~\ref{eq:y} is distinct
from the strong phase in the simulated quantities
$\cos\gamma\cos\delta$ (Eq.~\ref{eq:cosGammaCosDelta}) and
$\cos(\gamma\pm\delta)$ (Eqs.~\ref{eq:cosGammaPlusDelta}
and~\ref{eq:cosGammaMinusDelta}).  To distinguish these two strong phases, we
denote the one entering Eq.~\ref{eq:y} by $\delta_{t_P'}$.
Below, we verify that the simulated values of $\cos\gamma\cos\delta$ and
$\cos(\gamma\pm\delta)$ are insensitive to the choice of $\delta_{t_P'}$.
In the second method of constraining $\cos\gamma$, we simulate
Eq.~\ref{eq:cosGammaPhik} with $\delta_{t_P'}=\delta$.

\begin{table}[ht]
\begin{center}
\begin{tabular}{ccccc}
\hline\hline
Parameter & \multicolumn{3}{c}{References} & Value \\
\hline
${\cal F}$ & \multicolumn{3}{c}{\cite{Alexander:2000tb,Athar:2002mr}}&
	$1.11\pm 0.07$ \\
$f_{K^*}/f_\rho$ & \multicolumn{3}{c}{\cite{Chiang:2001ir}} & $1.04\pm 0.02$\\
$\left|V_{us}\right|$ & \multicolumn{3}{c}{\cite{pdg}} & $0.2196\pm 0.0026$ \\
$\left|V_{ud}\right|$ & \multicolumn{3}{c}{\cite{pdg}} & $0.9734\pm 0.0008$ \\
$\left|V_{ub}\right|$ & \multicolumn{3}{c}{\cite{pdg}} &
	$(3.66\pm 0.51)\cdot 10^{-3}$ \\
$\left|V_{cb}\right|$ & \multicolumn{3}{c}{\cite{pdg}} &
	$(4.07\pm 0.10)\cdot 10^{-2}$ \\
${\cal A}_{CP}(K^{*\pm}\pi^\mp)$ & \multicolumn{3}{c}{\cite{CLEOAcpKstpi}} &
	$0.26^{+0.33}_{-0.34}$$^{+0.10}_{-0.08}$ \\
$A$ & \multicolumn{3}{c}{\cite{Babar3pi}} & $-0.18\pm 0.08\pm 0.03$ \\
$C$ & \multicolumn{3}{c}{\cite{Babar3pi}} & $ 0.36\pm 0.18\pm 0.04$ \\
$\Delta C$ & \multicolumn{3}{c}{\cite{Babar3pi}} &
	$0.28^{+0.18}_{-0.19}\pm 0.04$ \\
$\rho_{AC}$ & \multicolumn{3}{c}{\cite{Hoecker}} & $-0.080$ \\
$\rho_{A\Delta C}$ & \multicolumn{3}{c}{\cite{Hoecker}} & $-0.059$ \\
$\rho_{C\Delta C}$ & \multicolumn{3}{c}{\cite{Hoecker}} & $0.176$ \\
\cline{2-4}
& CLEO & {\sc BaBar} & Belle \\
\cline{2-4}
${\cal B}(\rho^\pm\pi^\mp)$ & $27.6^{+8.4}_{-7.4}\pm 4.2$~\cite{CLEOPV}
              & $22.6\pm 1.8\pm 2.2$~\cite{Babar3pi}
              & $20.8^{+6.0}_{-6.3}$$^{+2.8}_{-3.1}$~\cite{BelleRhopi}
              & $22.8\pm 2.5$ \\
${\cal B}(^{^{(}}\!\bar K^{^{)}*0}\pi^\pm)$ &
	$7.6^{+3.5}_{-3.0}\pm 1.6$~\cite{CLEOPV}
              & $15.5\pm 1.8$$^{+1.5}_{-3.2}$~\cite{BabarKst0pi}
              & $19.3^{+4.2}_{-3.9}$$^{+4.1}_{-7.1}$~\cite{BelleKst0pi}
              & $12.4\pm 2.5$ \\
$\begin{array}{c} {\cal B}(K^{*\pm}_{K^0_S\pi^\pm}\pi^\mp) \\ {\cal B}(K^{*\pm}_{K^\pm\pi^0}\pi^\mp) \end{array}$ &
	$\biggl.\biggr\}16^{+6}_{-5}\pm 2$~\cite{CLEOKstpi}
              &
              & $\begin{array}{c} 20.3^{+7.5}_{-6.6}\pm 4.4\text{~\cite{BelleKstpi}}\\
		13.0^{+3.9+2.0+6.9}_{-3.6-1.8-6.1}\text{~\cite{BelleKstpi2}}\end{array}$
              & $\biggl.\biggr\}16.4^{+4.2}_{-4.0}$ \\
${\cal B}(\phi K^\pm)$    & $5.5^{+2.1}_{-1.8}\pm 0.6$~\cite{CLEOPhik}
              & $10.0^{+0.9}_{-0.8}\pm 0.5$~\cite{BabarPhik}
              & $10.7\pm 1.0^{+0.9}_{-1.6}$~\cite{BellePhik}
              & $9.6\pm 0.8$ \\
${\cal B}(\phi$$^{^{(}}\!\bar K^{^{)}0})$    &
	$5.4^{+3.7}_{-2.7}\pm 0.7$~\cite{CLEOPhik}
              & $7.6^{+1.3}_{-1.2}\pm 0.5$~\cite{BabarPhik}
              & $10.0^{+1.9}_{-1.7}$$^{+0.9}_{-1.3}$~\cite{BellePhik}
              & $8.1\pm 1.1$ \\
\hline
${\cal B}(\rho^\pm\pi^\mp)_P$ & \multicolumn{3}{c}{\cite{Babar3pi,Hoecker,CLEOPV,BelleRhopi}} &
	$13.9\pm 2.7$ \\
$\Delta(\rho^-\pi^+)$ & \multicolumn{3}{c}{\cite{Babar3pi,Hoecker,CLEOPV,BelleRhopi}} &
	$-2.9\pm 4.6$ \\
${\cal B}(K^{*+}\pi^-)$ & \multicolumn{3}{c}{\cite{Babar3pi,Hoecker,Chiang:2001ir,CLEOPV,BelleRhopi,CLEOAcpKstpi,CLEOKstpi,BelleKstpi,BelleKstpi2}} &
	$14.4^{+4.4}_{-4.0}$ \\
${\cal B}(K^{*-}\pi^+)$ & \multicolumn{3}{c}{\cite{Babar3pi,Hoecker,Chiang:2001ir,CLEOPV,BelleRhopi,CLEOAcpKstpi,CLEOKstpi,BelleKstpi,BelleKstpi2}} &
	$18.7^{+4.8}_{-4.6}$ \\
\hline\hline
\end{tabular}
\caption{Input parameters used to constrain $\gamma$.
Branching fractions and partial rate differences are given in units of
$10^{-6}$. Except for the last two entries, branching fractions are averaged
over charge conjugate states.}
\label{table:inputSummary}
\end{center}
\end{table}

Experimental measurements of the following quantities are given as input to
the simulation: ${\cal F}$, $f_{K^*}/f_\rho$, $\left|V_{us}\right|$, $\left|V_{ud}\right|$,
$\left|V_{ub}\right|$, $\left|V_{cb}\right|$, ${\cal A}_{CP}(K^{*\pm}\pi^\mp)$,
the $B\to\rho^\pm\pi^\mp$ parameters $A$, $C$, and $\Delta C$,
and the $CP$-averaged branching fractions for $B\to\rho^\pm\pi^\mp$,
$^{^{(}}\!\bar K^{^{)}*0}\pi^\pm$, $K^{*\pm}\pi^\mp$,
$\phi K^\pm$, and $\phi ^{^{(}}\!\bar K^{^{)}0}$.
These parameters are simulated with Gaussian or bifurcated Gaussian (different
widths above and below the peak) distributions, and their values are
summarized in Table~\ref{table:inputSummary}.
The input that contributes the largest uncertainty to $\gamma$ is the
$B\to K^{*\pm}\pi^\mp$ branching fraction.

For the five branching fractions, we combine all publicly presented
measurements, with statistical and systematic errors added in 
quadrature.  Where possible, the contribution from
$f_{+-}/f_{00}$ to the systematic error has been removed, since it is
included coherently in the simulation.  We neglect all other correlations
among the systematic errors.

\begin{figure}
\includegraphics*[width=5in]{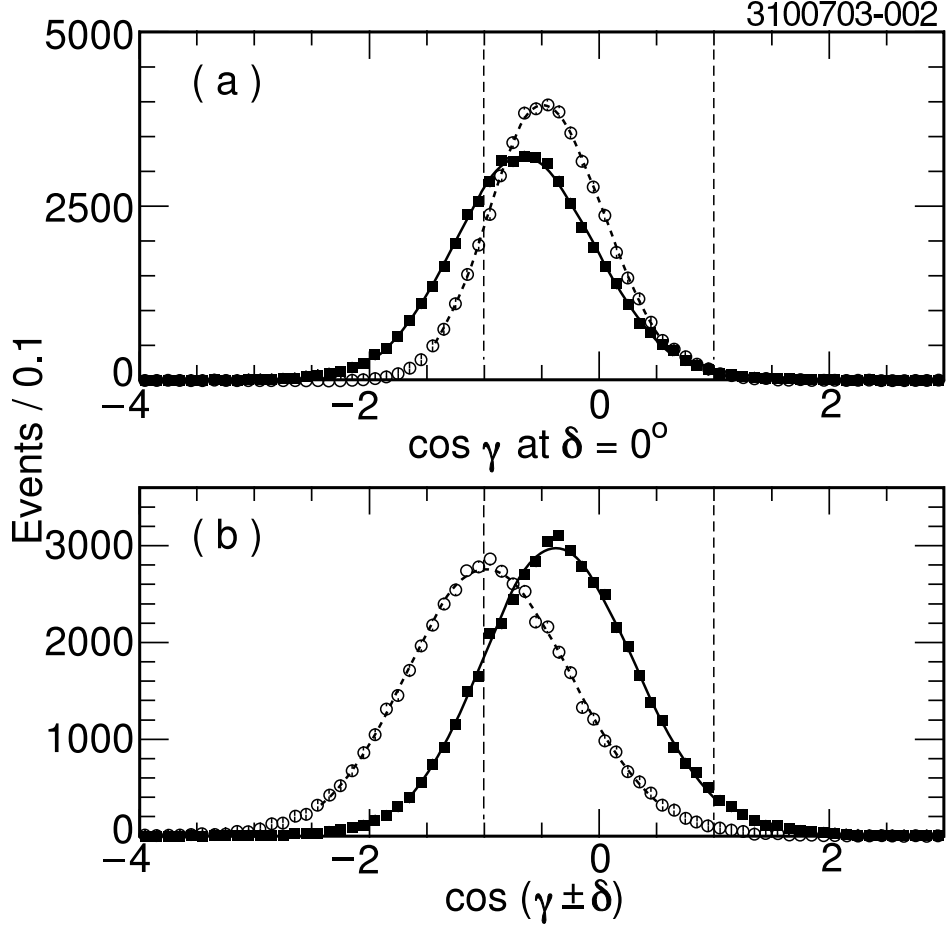}
\caption{Simulated distributions of $\cos\gamma\cos\delta$ from
Eq.~\ref{eq:cosGammaCosDelta} [(a), solid squares],
$\cos(\gamma + \delta)$ from Eq.~\ref{eq:cosGammaPlusDelta}
[(b), solid squares], and
$\cos(\gamma - \delta)$ from Eq.~\ref{eq:cosGammaMinusDelta}
[(b), open circles] using $B\to K^{*\pm}\pi^\mp$, as well as
$\cos\gamma$ from Eq.~\ref{eq:cosGammaPhik} [(a), open circles]
using $B\to K^{*\pm}\pi^\mp$ and $B\to\phi ^{^{(}}\!\bar K^{^{)}}$
with $|P_{\rm EW}^{\prime P}/p_P'|=0.5$, all with
$\delta_{t_P'}=0^\circ$. Overlaid on the histograms are the fits to bifurcated
Gaussians.  The dashed lines demarcate the physical region.}
\label{fig:cosGamma}
\end{figure}

Figure~\ref{fig:cosGamma} shows the simulated distribution of
$\cos\gamma\cos\delta$ from Eq.~\ref{eq:cosGammaCosDelta}, with
$\delta_{t_P'}=0^\circ$.
Fitting this distribution to a bifurcated Gaussian yields the measurement
$\cos\gamma\cos\delta = -0.68^{+0.63}_{-0.59}$, which suggests constructive
interference between $t_P'$ and $p_P'$.
The 90\%, 95\%, and 99\% C.L. upper limits on 
$\cos\gamma\cos\delta|_{\delta_{t_P'}=0^\circ}$ are 0.16, 0.42, and 0.94,
respectively.
Based on the smallness of direct $CP$ asymmetries in $B\to K\pi$, one can infer
a strong phase between tree and penguin amplitudes in these decays of
$(8\pm 10)^\circ$~\cite{Neubert:2002tf}.
If the strong phases in $B\to PV$ decays are as small as in two-pseudoscalar
($PP$) final states,
then our analysis weakly favors $\cos\gamma < 0$.
The variation of $\cos\gamma\cos\delta$ with $\cos\delta_{t_P'}$ is roughly
linear,
with a slope of $\frac{d\cos\gamma\cos\delta}{d\cos\delta_{t_P'}} = 0.11$.
Also shown in Figure~\ref{fig:cosGamma} is the distribution of
$\cos\gamma$ from Eq.~\ref{eq:cosGammaPhik}, with
$\delta=\delta_{t_P'}=0^\circ$ and
$|P_{\rm EW}^{\prime P}/p_P'|=0.5$.
Here, we obtain $\cos\gamma|_{\delta= 0^\circ} = -0.50^{+0.53}_{-0.47}$
and 90\%, 95\%, and 99\% C.L. upper limits on $\cos\gamma|_{\delta= 0^\circ}$
of 0.23, 0.44, and 0.89.

From Figure~\ref{fig:cosGamma}, we also find $\cos(\gamma + \delta)$ and
$\cos(\gamma - \delta)$ from Eqs.~\ref{eq:cosGammaPlusDelta}
and~\ref{eq:cosGammaMinusDelta} to be $-0.39^{+0.69}_{-0.63}$ and
$-0.99^{+0.74}_{-0.69}$, respectively, with a correlation coefficient of
$0.61$.  Considering only the 47\% of trials where
both quantities acquire physical
values, the distributions of the weak and strong phases imply
$\gamma = (113^{+20}_{-30})^\circ$ and $\delta=(-13\pm 17)^\circ$, with a
correlation coefficient of $7\cdot 10^{-5}$.
Because of the fourfold ambiguity of the $\gamma/\delta$ system, we fix
$\delta_{t_P'}$ to $0^\circ$ rather than equating it to the simulated value
of $\delta$.
The variations of $\cos(\gamma + \delta)$ and $\cos(\gamma - \delta)$ with
$\cos\delta_{t_P'}$ are given by
$\frac{d\cos(\gamma + \delta)}{d\cos\delta_{t_P'}}=0.06$
and $\frac{d\cos(\gamma - \delta)}{d\cos\delta_{t_P'}}=0.13$.  The values of
$\gamma$ and $\delta$ both change by less than $2^\circ$ between
$\delta_{t_P'}=0^\circ$ and $\delta_{t_P'}=80^\circ$.

\begin{figure}
\includegraphics*[width=5in]{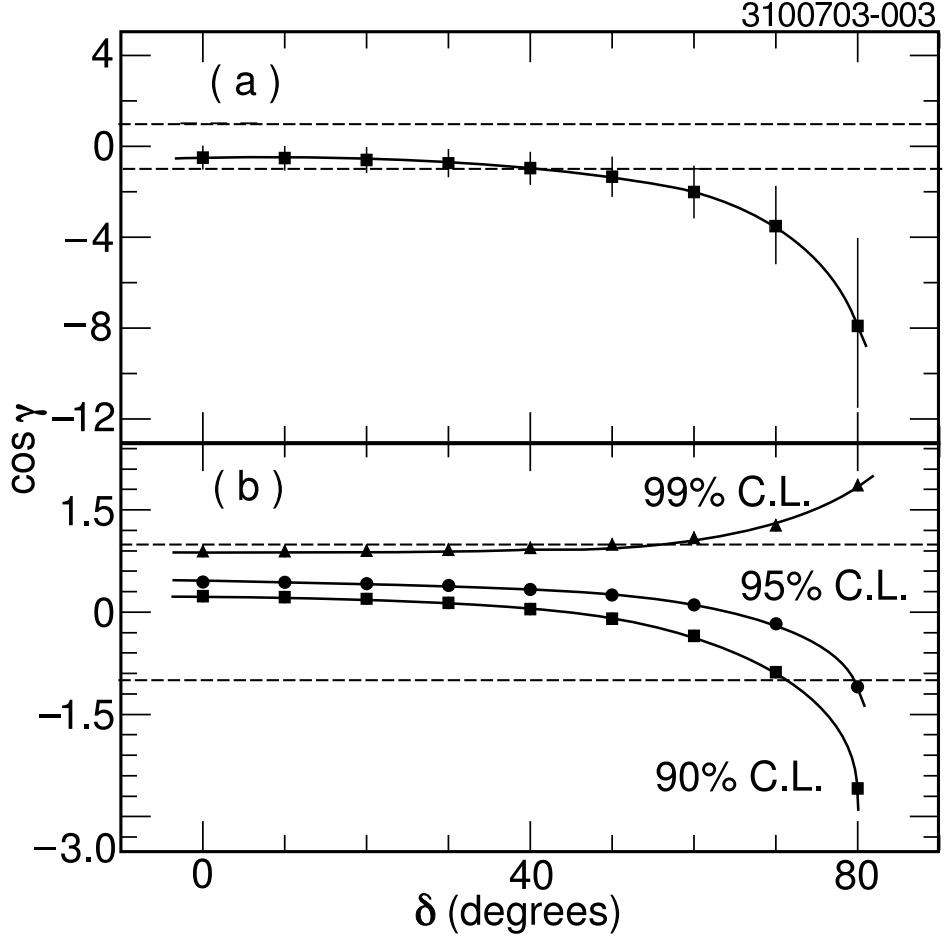}
\caption{Peak values (a) of and
upper limits (b) on $\cos\gamma$ from Eq.~\ref{eq:cosGammaPhik}
as a function of $\delta=\delta_{t_P'}$, using
$B\to K^{*\pm}\pi^\mp$ and $B\to\phi ^{^{(}}\!\bar K^{^{)}}$ with
$|P_{\rm EW}^{\prime P}/p_P'|=0.5$.  The asymmetric errors on the
peak values give the bifurcated Gaussian widths of the simulated distributions.
The dashed lines demarcate the physical region.}
\label{fig:cosGammaWithPhik}
\end{figure}

Figure~\ref{fig:cosGammaWithPhik} shows the dependence of
$\cos\gamma$ from Eq.~\ref{eq:cosGammaPhik} on $\delta=\delta_{t_P'}$,
with $|P_{\rm EW}^{\prime P}/p_P'|=0.5$.
The peak values are plotted with asymmetric error bars representing the widths
of the bifurcated Gaussian distributions.
By demanding that $\cos\gamma$ peak in the physical region, one can infer
that $\left|\delta\right|<41^\circ$.  The variation of $\cos\gamma$ with
$|P_{\rm EW}^{\prime P}/p_P'|$ is linear,
with a slope
$\frac{d\cos\gamma}{d\left|P_{\rm EW}^{\prime P}/p_P'\right|} = 0.28 - 1.51\cos\delta_{t_P'}$.
Incorporating the $B\to\phi ^{^{(}}\!\bar K^{^{)}}$ decays in
the measurement of $\gamma$
results in greater precision than using $B\to K^{*\pm}\pi^\mp$ alone,
but the theoretical uncertainties incurred are also larger.

Using the simulation of Eq.~\ref{eq:cosGammaCosDelta}, we also determine
the ratio $r=0.30^{+0.07}_{-0.05}$ at $\delta_{t_P'}=0^\circ$ with a
$\delta_{t_P'}$ dependence
given by $r = 0.25 + 0.09\cos\delta_{t_P'} - 0.04\cos^2\delta_{t_P'}$.
The inverse ratio for $\Delta S=0$ decays,
$\left|p_P/t_P\right| = \frac{1}{r}\left|V_{us}/V_{ud}\right|\left|V_{td}/V_{ts}\right|$,
is found to be $0.43-0.49\cos\delta_{t_P'}+0.26\cos^2\delta_{t_P'}$,
which takes the value $0.20^{+0.03}_{-0.02}$ at $\delta_{t_P'}=0^\circ$.

The widths of the generated distributions presented above are dominated by
experimental uncertainties on the input branching fractions,
${\cal A}_{CP}(K^{*\pm}\pi^\mp$), $A$, $C$, and $\Delta C$.
We study the improvement in the resolutions of
$\cos\gamma\cos\delta$, $\cos(\gamma\pm\delta)$, and
$\cos\gamma|_{\delta=0^\circ}$, collectively denoted by
$\hat\sigma_{\cos\gamma}$,
as these measurement uncertainties are reduced while maintaining the
central values at their current positions, with $\delta_{t_P'}=0^\circ$ and
$|P_{\rm EW}^{\prime P}/p_P'|=0.5$.  It is found that
$\hat\sigma_{\cos\gamma}$ scales with the size of the experimental
uncertainties until the latter reach 10\% of their current values, where
the resolution of $\gamma$ is ${\cal O}(10^\circ)$.
At this point, $\hat\sigma_{\cos\gamma}$ begins to be dominated by the
uncertainty on ${\cal F}$, and only by lowering $\sigma_{\cal F}$ can
$\hat\sigma_{\cos\gamma}$ be reduced any further.  

We have modeled SU(3) symmetry breaking effects in ratios of $\Delta S=0$ to
$\left|\Delta S\right|=1$ amplitudes with the purely real ratio of
decay constants $f_{K^*}/f_\rho$.  Repeating the simulation without SU(3)
breaking ({\it i.e.}, with $f_{K^*}/f_\rho=1$) results in changes to
$\cos\gamma\cos\delta$, $\cos(\gamma\pm\delta)$, and
$\cos\gamma|_{\delta=0^\circ}$ of 0.05 or smaller. Recent studies based on
QCD-improved factorization~\cite{Dariescu:2002hw,Beneke:2003zw} have suggested
that SU(3) breaking could be as large as 30\% and that the amplitude
ratios may possess a small complex phase.  To probe the impact of such effects,
we reinterpret $f_{K^*}/f_\rho$ as a phenomenological parameter and scale it
by $\pm 30\%$ of the value given in Table~\ref{table:inputSummary}, neglecting
any possible complex phases.  We find shifts of
$^{+0.21}_{-0.32}$ in $\cos\gamma\cos\delta$,
$^{+0.32}_{-0.45}$ in $\cos(\gamma+\delta)$,
$^{+0.12}_{-0.18}$ in $\cos(\gamma-\delta)$,
and $^{+0.19}_{-0.30}$ in $\cos\gamma|_{\delta=0^\circ}$.
Thus, in this conservative estimate, SU(3) breaking effects are roughly
15\%--70\% of the current experimental uncertainties.  To obtain meaningful
constraints on $\gamma$, future experimental advances must be accompanied by an
improved understanding of SU(3) breaking.

In conclusion, we have formed constraints on $\gamma$ as a function of
$\delta$ and $|P_{\rm EW}^{\prime P}/p_P'|$ using branching
fractions of and $CP$ asymmetries in $B\to PV$ decays.  At present,
experimental uncertainties overwhelm the theoretical uncertainties
arising from the model dependence of $\left|V_{ub}\right|$ and
$\left|V_{cb}\right|$, but they are the same order of magnitude as the
uncertainties in SU(3) symmetry breaking.
For strong phases of ${\cal O}(10^\circ)$ or smaller, our analysis favors
$\cos\gamma < 0$, which agrees with indications from $B\to PP$
decays~\cite{Neubert:2002tf,benekePPFit,fleischer-matias}.  However, the
current experimental precision does not yet permit a stringent comparision
with fits reliant upon $B$ and $K$ mixing.

\begin{acknowledgments}
We wish to thank Jonathan L.~Rosner and Alan J.~Weinstein for their
encouragement and perceptive guidance.
We are also grateful to Andreas H\"ocker and Yong-Yeon Keum
for helpful discussions.
This work was supported in part by the Department of Energy under Grant No.
DE-FG03-92ER40701.
\end{acknowledgments}

\end{document}